\documentclass[twocolumn]{aastex702}

\usepackage[T1]{fontenc}
\usepackage{ae,aecompl}

\usepackage[normalem]{ulem} 

\usepackage{graphics,epsf}
\usepackage[utf8]{inputenc}
\usepackage{amsmath}                
\usepackage{amsfonts}               
\usepackage{amssymb}                
\usepackage{epsfig}                 
\usepackage{graphicx}
\usepackage{float}
\usepackage{color}
\usepackage{multirow}               
\usepackage{hyperref}
\usepackage{xspace}
\usepackage{rotating}

\hypersetup{
    colorlinks=true,
    linkcolor=red,   
    urlcolor=cyan}

\usepackage[para,online,flushleft]{threeparttable}
\usepackage{lineno}

\newcommand{\cm}{{~\rm cm}}
\newcommand{\km}{{~\rm km}}
\newcommand{\s}{{~\rm s}}

\newcommand{\g}{{~\rm g}}
\newcommand{\G}{{~\rm G}}
\newcommand{\K}{{~\rm K}}
\newcommand{\erg}{{~\rm erg}}
\newcommand{\yr}{{~\rm yr}}

\newcommand{\AU}{{~\rm AU}}

\shorttitle{Wobbling jets in CEE}
\shortauthors{Hillel, Schreier, Soker}

\begin{document}

\title{Simulating the convection in red supergiant stars: wobbling jets in common envelope evolution}

\author{Shlomi Hillel}
\affiliation{Department of Physics, Technion - Israel Institute of Technology, Haifa, 3200003, Israel; shlomi.hillel@gmail.com, ronsr@technion.ac.il, soker@technion.ac.il}
\email{shlomi.hillel@gmail.com}

\author{Ron Schreier}
\affiliation{Department of Physics, Technion - Israel Institute of Technology, Haifa, 3200003, Israel; shlomi.hillel@gmail.com, ronsr@technion.ac.il, soker@technion.ac.il}
\email{ronsr@technion.ac.il}

\author[0000-0003-0375-8987]{Noam Soker} 
\affiliation{Department of Physics, Technion - Israel Institute of Technology, Haifa, 3200003, Israel; shlomi.hillel@gmail.com, ronsr@technion.ac.il, soker@technion.ac.il}
\email{soker@technion.ac.il}
\correspondingauthor{Noam Soker: soker@technion.ac.il}

\begin{abstract}
We use our newly constructed three-dimensional red supergiant (RSG) stellar model, which also mimics nuclear energy production and photospheric emission, to calculate the stochastic component of the angular momentum of the mass that a companion spiraling within the RSG's envelope accretes during common envelope evolution (CEE). The accreted mass has a fixed-direction angular-momentum component arising from the density gradient in the RSG envelope and orbital motion. The angular momentum component with a stochastically varying direction results from vigorous envelope convection. We do not include the companion's influence on the RSG envelope during the CEE and consider an undisturbed, non-rotating RSG stellar model. We find that the fluctuating angular momentum amplitude can be several times the fixed-axis angular momentum. The total specific angular momentum of the accreted mass easily forms intermittent accretion disks around neutron stars and black holes, but it is only marginally sufficient, or not at all, to form accretion disks around main-sequence stellar companions. The intermittent accretion disks we expect to form will launch wobbling jets with varying axes. We discuss aspects of wobbling jets in the CEE and the grazing envelope evolution (GEE), which might precede the CEE or replace it altogether. Studies have claimed that jets are a crucial ingredient in many cases of CEE, and the standard CEE should include jets that the companion launches, before (like the GEE), during, and/or at the exit from the CEE. Our study supports this claim and emphasizes the importance of wobbling jets. 
\end{abstract} 


\section{Introduction}
\label{sec:Introduction}

In the common envelope evolution (CEE; e.g., \citealt{RoepkeDeMarco2023} for a recent review, and \citealt{Borges2026, DiStefanoetal2026, Gurjaletal2026, Karinoetal2026, LiZetal2026, MuCetal2026, Noughanietal2026, ShiYetal2026, Thaietal2026, Yamaguchietal2026} for some recent CEE studies), a smaller star spirals into the envelope of a giant star and removes part or all of the giant's envelope. The companion can survive in a binary system with the core of the giant star or interact with the giant's core. In the case of core-companion interaction, either they merge, or one of them destroys the other. In the traditional CEE (e.g., \citealt{Paczynski1976, Webbink1984}), the only energy available to remove the giant's envelope in a short time is the orbital energy of the companion-core system. However, over the years, it became clear that in many cases the companion accretes mass (e.g. \citealt{Jonesetal2015}) and launches jets (e.g., \citealt{Grichener2025}). It seems that the presence of jets in the CEE is the most robust observable \citep{Soker2025Robust}, besides the close binary system of a remnant (a white dwarf, a neutron star [NS], or a black hole) and a companion that testifies to the CEE.
In our study of the CEE (e.g., \citealt{Hilleletal2023, Schreieretal2025, Soker2025WDCEE, WeinerSoker2025}), we consider the present standard CEE to include jets that the companion launches. Namely, \textit{the standard CEE is not the traditional CEE}.   

The argument for the major role of jets in CEE primarily comes from planetary nebulae. Many planetary nebulae exhibit axisymmetric, i.e., one symmetry axis, or multipolar, i.e., two or more symmetry axes, morphologies (e.g., \citealt{Balick1987, Chuetal1987, Schwarzetal1992, CorradiSchwarz1995, SahaiTrauger1998, Sahaietal2007, Sahaietal2011, Parkeretal2016, Parker2022}, for papers and catalogs with large collections of images). 
More than a hundred post-asymptotic giant branch nebulae and planetary nebulae have central binary systems (e.g., \citealt{Miszalski2019ic, Oroszetal2019, Jones2020Galax, Jones2025}); these are post-CEE binary systems, or post-grazing envelope evolution (GEE) binary systems. Some main-sequence companions are inflated, indicating that they have accreted mass in the CEE (e.g., \citealt{Jonesetal2015}).
Many studies consider pairs of jet to shape these pairs of structural features, e.g.,  pairs of dense clumps, ears, or lobes (e.g., \citealt{Morris1987, Soker1990AJ, GarciaSegura1997, SahaiTrauger1998, GarciaSeguraLopez2000, GarciaSeguraetal2005, BlackmanLucchini2014, Tocknelletal2014, GarciaSeguraetal2021, GarciaSeguraetal2022, AkashiSoker2018, EstrellaTrujilloetal2019, Tafoyaetal2019, Balicketal2020, RechyGarciaetal2020, Clairmontetal2022, Danehkar2022, MoragaBaezetal2023, Derlopaetal2024, Mirandaetal2024, Sahaietal2024}; note that \citealt{Baanetal2021} discussed an alternative shaping mechanism). The post-CEE (or post-GEE) binary systems at the center of jet-shaped nebulae strongly hint at the major roles of jets in CEE \citep{Soker2025Robust}. 

Most hydrodynamical simulations of the CEE do not include jets (e.g., \citealt{Staffetal2016MN, Kuruwitaetal2016, Ohlmannetal2016a,  Iaconietal2017b, Chamandyetal2019, LawSmithetal2020, GlanzPerets2021a, GlanzPerets2021b, GonzalezBolivaretal2022, GonzalezBolivaretal2024, Lauetal2022a, Lauetal2022b,  BermudezBustamanteetal2024, Chamandyetal2024, GagnierPejcha2024,  Landrietal2024, RosselliCalderon2024, Vetteretal2024, Bhattacharyyaetal2025, Vetteretal2025, Gagnieretal2026}, for some papers from the last decade); some do not even acknowledge the major roles of jets. 
A minority of CEE and GEE simulations do include jets that the companion launches, but pay the price in ignoring some other effects, like the gravity of the companion or the accretion process, or the simulation lasts for a short time (e.g., \citealt{MorenoMendezetal2017, ShiberSoker2018, LopezCamaraetal2019, Shiberetal2019, LopezCamaraetal2020MN, Hilleletal2022, Hilleletal2023, LopezCamaraetal2022, Zouetal2022, Soker2022Rev, Zouetal2022, Schreieretal2023, Schreieretal2025, Gurjareta2024eas, ShiberIaconi2024, Gurjaletal2026}). The interaction of the jets with the common envelope and the accretion processes that launch the jets are part of a feedback cycle (e.g., \citealt{Soker2016Rev, GrichenerCohenSoker2021, Hilleletal2022, WeinerSoker2025, Gurjaletal2026}), which is hard to simulate in full. 

To launch a jet, the accretion onto the companion should be via an accretion disk (or possibly an accretion belt, e.g., \citealt{SchreierSoker2016}), implying that the accreted mass has sufficiently high specific angular momentum.  There are two sources of angular momentum in cases of a single companion (in the case of a tight binary system that enters the CEE, the tight binary system can supply angular momentum, e.g., \citealt{Schreieretal2019}). The first is the orbital motion of the companion within the envelope, which has a radial density gradient. This angular momentum component has a fixed axis aligned with the orbital angular momentum. The second is the envelope convection, which introduces a stochastic component to the angular momentum of the accreted mass (e.g., \citealt{Dorietal2023}). We elaborate on these components in Section \ref{sec:AMoMethod}. 
  
In this study, we use a three-dimensional (3D) simulation of a single RSG star from \cite{Schreieretal2026} to estimate the relative contribution of the stochastic angular-momentum component due to convection in the envelope; the contribution is to the mass that a neutron star (NS), a black hole, or a main-sequence companion accretes while orbiting the outer envelope of an RSG star. We describe the hydrodynamical simulation in Section \ref{sec:Methods}. In Section \ref{sec:AMoMethod}, we present the expressions for the angular momentum due to the orbital motion and due to the envelope convection. In Section \ref{sec:AMoResults}, we present the results of our application of these expressions to the stellar model. 
There are other significant effects of envelope convection that we will not study here, although they are relevant to the CEE. Most importantly, convection can efficiently transfer energy to the photosphere (e.g., \citealt{Sabachetal2017, WilsonNordhaus2019}), thereby reducing the efficiency of envelope removal (e.g., \citealt{WilsonNordhaus2020, WilsonNordhaus2022}).
In Section \ref{sec:Summary}, we summarize how this study advances our understanding of the role of jets in standard CEE.

\section{The RSG stellar model}
\label{sec:Methods}

For our calculations, we use simulations from \cite{Schreieretal2026}, where we included two new numerical ingredients to those of earlier papers (e.g., \citealt{Hilleletal2023, Schreieretal2025}): (1) We mimicked the emission from the photosphere of the RSG by cooling each grid cell with a density below about the initial photospheric density $2.1 \times 10^{-9} \g \cm^{-1}$. Specifically, in each grid cell with a density of 
$\rho < \rho_{\rm ph}=2\times 10^{-9} \g \cm^{-3}$, we set the temperature to be $T_{\rm CSM}= 1000 \K$, much lower than the photospheric temperature of $T_{\rm ph}=3160 \K$. 
(2) To mimic the nuclear energy production in the core, we deposited energy with a power equal to the luminosity of the 1D stellar model, $L= 2.65 \times 10^{38} \erg \ s^{-1} \simeq 7 \times 10^4 L_\odot$, into a shell bounded by $14 \times 10^{12} \cm = 201 R_\odot < r  < 16 \times 10^{12} \cm = 230 R_\odot $, which is above and close to the inert core.
In what follows, `energy deposition' refers to the energy that we deposit to mimic the nuclear burning in the RSG core. 

To substantially reduce computational time, we do not follow the evolution of an inner inert core, i.e., a sphere with a radius of $R_{\rm inert} = 0.2R_{\rm RSG} = 176 R_{\odot}$ and a mass of $5.64 M_\odot$. The gravitational force in the simulation matches that in the 1D code when we transfer it to the 3D grid, including the gravity of the inert core. 
The inert core is a large central sphere that extends into the envelope, which the numerical code does not follow. Namely, the simulations include only the gravity of this inert core.  
 
The 1D RSG stellar model comes from a zero-age-main-sequence star of mass $M_{\rm 1,ZAMS}=15 M_\odot$ and metallicity $Z=0.02$, which was evolved with the \texttt{MESA} stellar evolution code \citep{Paxtonetal2011, Paxtonetal2013, Paxtonetal2015, Paxtonetal2018, Paxtonetal2019}. When we transported this 1D RSG stellar model to the 3D grid, it had a mass of $M_1=12.5 M_\odot$ and a radius of $R_{\rm RSG}=881 R_{\odot} = 6.1 \times 10^{13} \cm$. 
We used the hydrodynamical numerical code {\sc flash} \citep{Fryxelletal2000} with a Cartesian grid and assumed the gas is an ideal gas with an adiabatic index of $\gamma = 5/3$, including radiation pressure. There is an uniform cell size of $\Delta _{\rm c}=L_{\rm G}/256 = 9.766 \times 10^{11} \cm$, for $L_{\rm G} = 250 \times 10^{12} \cm$. The center of the RSG is at the origin. 

To prevent the influence of the cubical boundary of the numerical grid, we \citep{Schreieretal2026}  imposed a numerical boundary on a sphere of radius $R_{\rm w}=100 \times 10^{12} \cm = 1437 R_\odot= 1.63 R_{\ast,0}$, where $R_{\ast,0}$ is the initial radius of the 1D stellar model. From this radius outward, all velocities are set to be radial outward with a velocity of $v=0.1 \km \s^{-1}$. 

We found that these two ingredients are necessary for stellar pulsation to continue; without them, pulsation decays (see \citealt{Schreieretal2026} for all details). 
  
\section{The sources of accreted angular momentum}
\label{sec:AMoMethod}

\cite{Dorietal2023} estimated the contribution of the two angular momentum sources to the accreted mass in the envelope of an RSG star, using a 1D stellar evolution model. They removed mass at a high rate and injected energy into the envelope to mimic the energy of the jets that the companion launches. We will differ by using a 3D model and by neglecting the energy that the companion injects into the envelope. Namely, we will examine the contribution of convection in the unperturbed envelope. This is because in the 3D simulations we will incorporate jet energy self-consistently by allowing changes in envelope structure to influence the accretion rate, and hence the jet power. This will require much more extensive simulations. We therefore postpone the injection of energy via jets to a later study, which is much more computationally intensive in 3D simulations. We present the main derivations from \cite{Dorietal2023} relevant to us. 

\subsection{Fixed-axis angular momentum by orbital motion}
\label{subsec:OrbitAM}

We consider a companion star of mass $M_2$ orbiting at about the Keplerian velocity at orbital radius $a$ within the envelope of an RSG star, accreting mass via Bondi-Hoyle-Lyttleton (BHL) type accretion at a rate below the classical rate \citep{Soker2004AM}. The relative companion-to-envelope velocity is 
\begin{equation}
v_r \simeq v_{\rm Kep}= \left[ \frac{G M_1(a)}{a} \right]^{1/2}, 
\label{eq:Vkep}
\end{equation}
where $M_1(a)$ is the giant mass inside radius $r=a$. The BHL accretion radius we use here is defined by the orbital motion. Considering the sound speed, this radius would be smaller for supersonic velocity (for accretion in a subsonic flow, see, e.g., \citealt{Gruzinov2022, Prustetal2024}). On the other hand, envelope rotation reduces the relative velocity and increases the accretion radius. For the present study, the following definition is adequate \citep{Dorietal2023}
\begin{equation}
\begin{split}
R_{\rm BHL} & \equiv \frac{2GM_2}{v^2_r} \simeq 0.4 
 \\ & \times 
\left( \frac{M_2}{0.2 M_1(a)} \right) 
\left( \frac{a}{1 \AU} \right) \AU. 
\label{eq:RBHL}
\end{split}
\end{equation}

Because of the orbital motion in the azimuthal direction and the density gradient in the radial direction, the envelope gas that the secondary star accretes has a non-zero angular momentum $j_{\rm O}$ (`O' stands for orbital).  
Like some earlier studies (e.g., \citealt{Soker2004AM, Dorietal2023}), we assume that the 3D hydrodynamical simulations of the BHL accretion of a point mass moving linearly in a box by \cite{Livioetal1986} hold for the CEE. The much more sophisticated 3D simulations of accretion from a wind performed by \cite{Kashietal2022} give us confidence in this assumption.  

For a density gradient perpendicular to the relative velocity of $d \rho /dr = -\rho (r)/H$, 
\cite{Livioetal1986} find the specific angular momentum of the accreted gas to be   
\begin{equation}
j_{\rm O}=\frac{\eta}{4 H} 
\frac{ (2 G M_2)^2}{v^3_r} = \eta \beta \left[ \frac{M_2}{M_1(a)} \right]^{2} a v_{\rm Kep} ,
\label{eq:jaccO1}
\end{equation}
with a typical value of $\eta \simeq 0.25$ in their simulations. In a CEE, the relative velocity of the material at larger radii than the orbit of the mass-accreting companion is larger than the velocity of the accreted material from smaller radii. This velocity asymmetry introduces angular momentum with an opposite direction to that from the density gradient. Therefore, we take an even lower value of $\eta$ and will scale with $\eta=0.1$. This value of $\eta$ serves as a scaling factor, as we do not know its exact value. It may be somewhat larger if the effect of the velocity asymmetry is small, or somewhat smaller.     
In the second equality of equation (\ref{eq:jaccO1}) we used equations (\ref{eq:Vkep}) and an envelope density profile of 
\begin{equation}
\rho_{\rm env} (r) \propto r^{-\beta}.
\label{eq:EnvDensty}
\end{equation}
  
This angular momentum component of the accreted gas has a constant direction perpendicular to the orbital plane. 
The radius of an accretion disk with the specific angular momentum from equation (\ref{eq:jaccO1}) is 
\begin{equation}
\begin{split}
 R_{\rm d2} = & \frac {j^2_{\rm O}}{GM_2} =\eta^2 \beta^2
 \left[ \frac{M_2}{M_1(a)} \right]^{3}  a = 0.155
\\ \times 
\left( \frac{\eta}{0.1} \right)^2  
&
\left( \frac{\beta}{3} \right)^2 
\left( \frac{M_2}{0.2 M_1(a)} \right)^3 
\left( \frac{a}{1 \AU} \right) R_\odot,
\label{eq:Rdisk2}
\end{split}
\end{equation}
where we used equation (\ref{eq:Vkep}), and the subscript d2 refers to a disk around the companion. 
The accretion process robustly forms an accretion disk around NSs and black holes because the accretion disk radius is much larger than their radii. It is marginal around main-sequence stars; although the radii of main-sequence stars are larger than $R_{\rm d2}$, as the scaling of equation (\ref{eq:Rdisk2}) gives, some favorable parameters like larger values of $\eta$ and $\beta$ can make $R_{\rm d2}$ larger. For example, in the outer envelope the density gradient is large, and $\beta \gg 1$. The scenario with main-sequence stars is that they form an accretion disk before entering the envelope, and this disk continues to survive as they spiral in towards the center (see papers on the GEE cited in Section \ref{sec:Introduction}).

\subsection{Stochastic angular momentum from convection}
\label{subsec:Stochastic}

We follow \cite{Dorietal2023} in the basic arguments, but substantially differ from them in the properties of the convective blobs. While they use mixing-length theory, we will use 3D simulations. For example, they took the number of convective blobs within a sphere of radius $R_{\rm BHL}$ to be $N_{\rm BHL} \simeq (R_{\rm BHL}/\lambda)^3$, where $\lambda$ is the mixing length. We, instead, will sum over the stochastic angular momentum in the sphere. 
The convective blobs in this sphere are accreted during a timescale of 
\begin{equation}
\begin{split}
\tau_{\rm BHL} & \simeq \frac{R_{\rm BHL}}{v_r} =
7.35 
\left( \frac{R_{\rm BHL}}{0.4 \AU} \right) 
\\ & \times 
\left( \frac{M_1(a)}{10 M_\odot} \right)^{-1/2} 
\left( \frac{a}{1 \AU} \right)^{1/2} ~{\rm day} . 
\label{eq:TBHL}
\end{split}
\end{equation}
  
The relevant timescale for determining the volume from which we should sum the stochastic angular momentum is the typical lifetime of the accretion disk, $\tau_{\rm d2}$; this volume is $(v_r \tau_{\rm d2})^3$. 
The radius of the disk is $R_{\rm d2}$ as given by equation (\ref{eq:Rdisk2}). As in \cite{Dorietal2023}, we take the typical disk's lifetime to be tens of times the Keplerian orbital period at that radius, $\zeta$. The disk's lifetime is 
\begin{equation}
\begin{split}
\tau_{\rm d2} &= \zeta \frac{ 2 \pi R_{d2}^{3/2}}{(G M_2)^{1/2}}  = 8.2
\left( \frac{\zeta}{100} \right)   
\\ &    \times
\left( \frac{M_2}{2 M_\odot} \right)^{-1/2}   
\left( \frac{R_{\rm d2}}{1 R_\odot} \right)^{3/2}   ~{\rm day} . 
\label{eq:taudiskD}
\end{split}
\end{equation}
We find that for the typical parameters we use here $\tau_{\rm BHL} \simeq \tau_{\rm d2}$. 
For that, we will estimate the specific angular momentum of the stochastic (random) component, $j_{\rm R}$, by simply summing over all numerical grid points within a spherical volume centered on the companion's orbit; the center of the volume is at points along the companion's orbit, and its radius $R_{\rm S}$ is a fraction of the BHL radius, $0.5R_{\rm BHL} \lesssim R_{\rm S} \le R_{\rm BHL}$.  The specific angular momentum of the stochastic component due to the random convection motion in this volume is 
\begin{equation}
\begin{split}
\vec{j}_{\rm R} = \frac{1}{M_{\rm S}} \Sigma V_i \rho_i \vec{r}_{i2} \times \vec{v}_i
\quad {\rm for} \quad {r}_{i2}<R_{\rm S}, 
\label{eq:jR1}
\end{split}
\end{equation}
where $\vec{r}_{i2} \equiv \vec{r}_i - \vec{r}_2$, $\vec{r}_i$ is the location of grid point $i$, $V_i$ its volume, $\rho_i$ its density, $\vec{r}_2$ is the location of the mass-accreting companion, and $M_{\rm S}$ the total mass in the sphere.

\begin{figure} 
\centering
\includegraphics[trim=0.0cm 16.0cm 0.0cm 2.3cm ,clip, angle=0, scale=0.74]{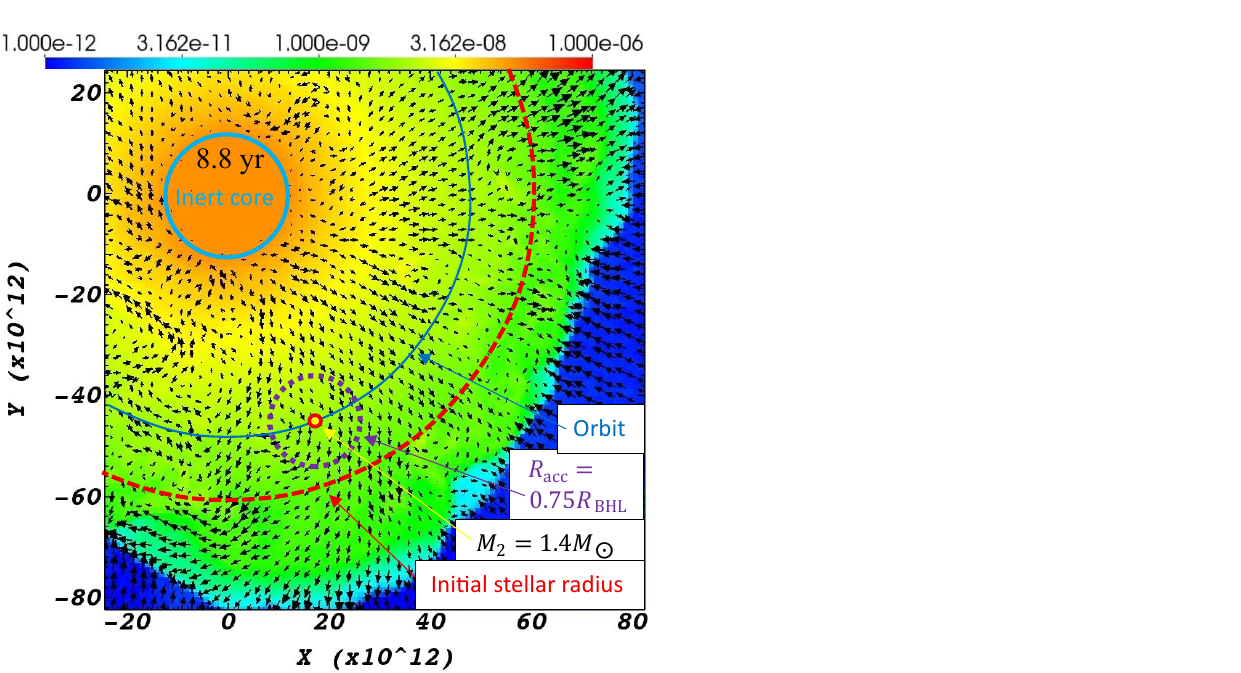}
\caption{The density and velocity map in the orbital plane $z=0$ that we use in this study for cases with an RSG stellar model with energy deposition that mimics the nuclear energy production of the star, for the time $t=8.8 \yr$ (from \citealt{Schreieretal2026}). The pale-blue circle is the inert core, and the red-dashed line is the initial surface of the giant: $R_{\ast,0}=881 R_\odot = 61 \times 10^{12} \cm$. The arrows depict the velocities: direction and magnitude relative to the arrow length. The maximum velocity is $71 \km \s^{-1}$. The colors depict the density according to the color bar in units of $\g \cm^{-3}$. The solid blue line depicts the circular orbit of $a=700 R_\odot$, and the dotted purple circle depicts a sphere of radius $R_{\rm acc}=0.75R_{\rm BHL}=130R_\odot$ centered on the temporary location of the companion of mass $M_2=1.4 M_\odot$. Inside this sphere, we calculate the stochastic angular-momentum component arising from convective motion. The clear radial density gradient is responsible for the constant-axis component of angular momentum (the axis perpendicular to the orbital plane).    
}
\label{fig:a700Schematic}
\end{figure}

Figure \ref{fig:a700Schematic} presents the flow and density maps in the equatorial plane at $t=8.8 \yr$ for the case with energy deposition, i.e., energy deposition to mimic the nuclear burning in the RSG core (for more details see \citealt{Schreieretal2026}). We also mark the orbit of $a=700 R_\odot$ (solid blue line). The dotted purple line depicts a circle with a radius of $R_{\rm acc}=0.75 R_{\rm BHL}$ around a companion of mass $M_2=1.4 M_\odot$. 

Figure \ref{fig:a700Schematic} shows that a typical radius of the convective vortices is $\simeq 10^{13}\cm$, which is about ten times the cell size. Therefore, our grid can resolve the convection eddies (vortices). We also note that the accretion radius might turn larger than some convective vortices. This is not a problem, as we do not demand that the accretion radius be larger than the size of vortices.

\section{The wobbling magnitude}
\label{sec:AMoResults}

We present the numerical results for a companion orbiting in the outer envelope, as we cannot calculate too close to the inert core. Specifically, we will present results for two sampled orbits: $a=450 R_\odot$ and $a=700 R_\odot$. 
We found that at these orbital radii, it is adequate to take the density scale height to be $H \simeq 200 R_\odot$, i.e., $\rho/(d \rho/dr) \simeq 200 R_\odot$ (which will not be accurate at smaller orbital radii). 
With this, we will take the constant orbital specific angular momentum from equation (\ref{eq:jaccO1}), the value of  
\begin{eqnarray}
\begin{split}
j_{\rm O} & = \eta \left[ \frac{M_2}{M_1(a)} \right]^{2} \frac{a^2}{H} v_{\rm Kep} = 1.05 \times 10^{18} \\ &  
\times \left( \frac{\eta}{0.1} \right)
       \left( \frac{M_2}{1.4M_\odot} \right)^2
       \left( \frac{M_1(a)}{10 M_\odot} \right)^{-3/2}   \\ & 
\times \left( \frac{a}{500 R_\odot} \right)^{3/2}
       \left( \frac{H}{200 R_\odot} \right)^{-1}
\cm^2 \s^{-1}. 
\label{eq:jaccO2}
\end{split}
\end{eqnarray}

The total specific angular momentum of the accretion inflow is  
\begin{equation}
\vec{j}_{\rm acc} = \vec j_{\rm O} + \vec{j}_{\rm R}.   
\label{eq:jacc}
\end{equation}
The component of the fixed-axis angular momentum is basically along the orbital angular momentum. The relative direction of the random component of angular momentum relative to the orbital angular momentum is given by 
\begin{equation}
\cos  \phi = \frac{ \vec j_{\rm O} \cdot \vec{j}_{\rm R} }{j_{\rm O}j_{\rm R}}. 
\label{eq:JRRand}
\end{equation}
The relative angle of the combined angular momentum relative to the orbital angular momentum is given by 
\begin{equation}
\cos  \alpha_{\rm w} = \frac{ \vec j_{\rm O} \cdot \vec{j}_{\rm acc} }{j_{\rm O}j_{\rm acc}}. 
\label{eq:JaccRand}
\end{equation}

We present results for a companion mass of $M_{\rm NS} = 1.4 M_\odot$, aiming particularly at an NS companion, as in common-envelope jets supernovae (CEJSNe), where an NS (and sometimes a black hole) accretes mass as it spirals in inside the envelope and then inside the core (e.g., \citealt{SokerGrichenerGilkis2019}), and CEJSN impostors (where the NS does not enter the core, e.g., \citealt{GilkisSokerKashi2019}; for a recent study of the accretion process into an NS, see \citealt{ChiKitCheongetal2026}). We consider an accretion radius of $R_{\rm acc}=0.75 R_{\rm BHL}$, where $R_{\rm BHL}$ is by equation (\ref{eq:RBHL}) for $M_2=1.4 M_\odot$. We assume that equation (\ref{eq:jaccO2}), with the scaling of low $\eta$ and the other variables, holds for $R_{\rm acc}=0.75 R_{\rm BHL}$.   
The goal of the figures to follow is to show the large fluctuations in the specific angular momentum of the accreted gas due to envelope convection.  

\begin{figure} 
\centering
\includegraphics[trim=0.6cm 0.5cm 0.0cm 0.0cm ,clip, angle=0, scale=0.48]{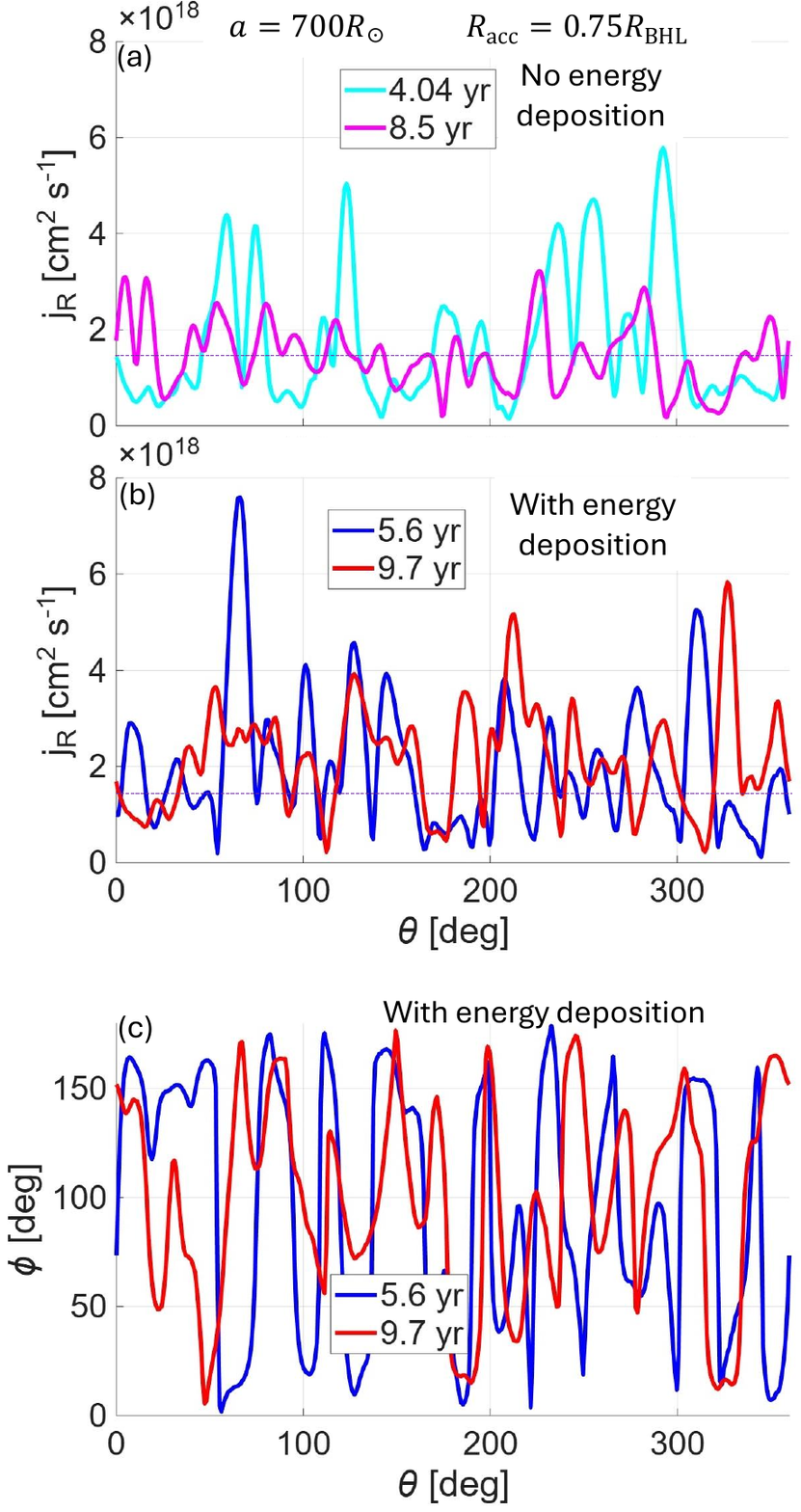}
\caption{The wobbling for a circular orbit at $a=700 R_\odot$. (a) The random specific angular momentum of the accreted mass due to the envelope convection according to equation (\ref{eq:jR1}) inside a sphere of radius $R_{\rm acc}=0.75 R_{\rm BHL}$, as a function of orbital angle, for the simulation with no energy deposition. The purple-dashed horizontal line marks $j_{\rm O}$ according to equation (\ref{eq:jaccO2}), but for $a=700 R_\odot$ and $M_1(a)=11.3 M_\odot$. (b). Similar to panel (a), but for the simulation with energy deposition that mimics the stellar energy production. The fluctuations are larger. (c) The angle of the accreted angular momentum relative to the orbital angular momentum as given by equation (\ref{eq:JRRand}) for the case shown in panel (b).  
}
\label{fig:a700}
\end{figure}

In Figure \ref{fig:a700} we present results at two times for a circular orbit at $a=700 R_\odot$ where $M_a(a)=11.3 M_\odot$. By equation (\ref{eq:jaccO2}), we find the specific angular momentum due to the orbital motion and the density gradients to be $j_{\rm O}(700)=1.45 \times 10^{18} \cm^2 \s^{-1}$, with the other parameters as in equation (\ref{eq:jaccO2}). We mark this value with a purple dashed horizontal line in panels (a) and (b), where we present $j_{\rm R}$ as a function of orbital angle along a whole orbit for the simulation with no energy deposition and with energy deposition, respectively. In panel (c), we present the angle of $\vec{j}_{\rm R}$ relative to the orbital angular momentum, $\phi$, as given by equation (\ref{eq:JRRand}). At late times, $t \gtrsim 8 - 10 \yr$, the fluctuations of the random component of the accreted angular momentum ${j}_{\rm R}$ are larger in the simulation with energy deposition. This is expected, as the convection velocity is larger in this simulation \citep{Schreieretal2026}. As we consider this simulation more realistic, we will present results only for this simulation.  

\begin{figure} 
\centering
\includegraphics[trim=0.0cm 9.8cm 0.0cm 0.0cm ,clip, angle=0, scale=0.48]{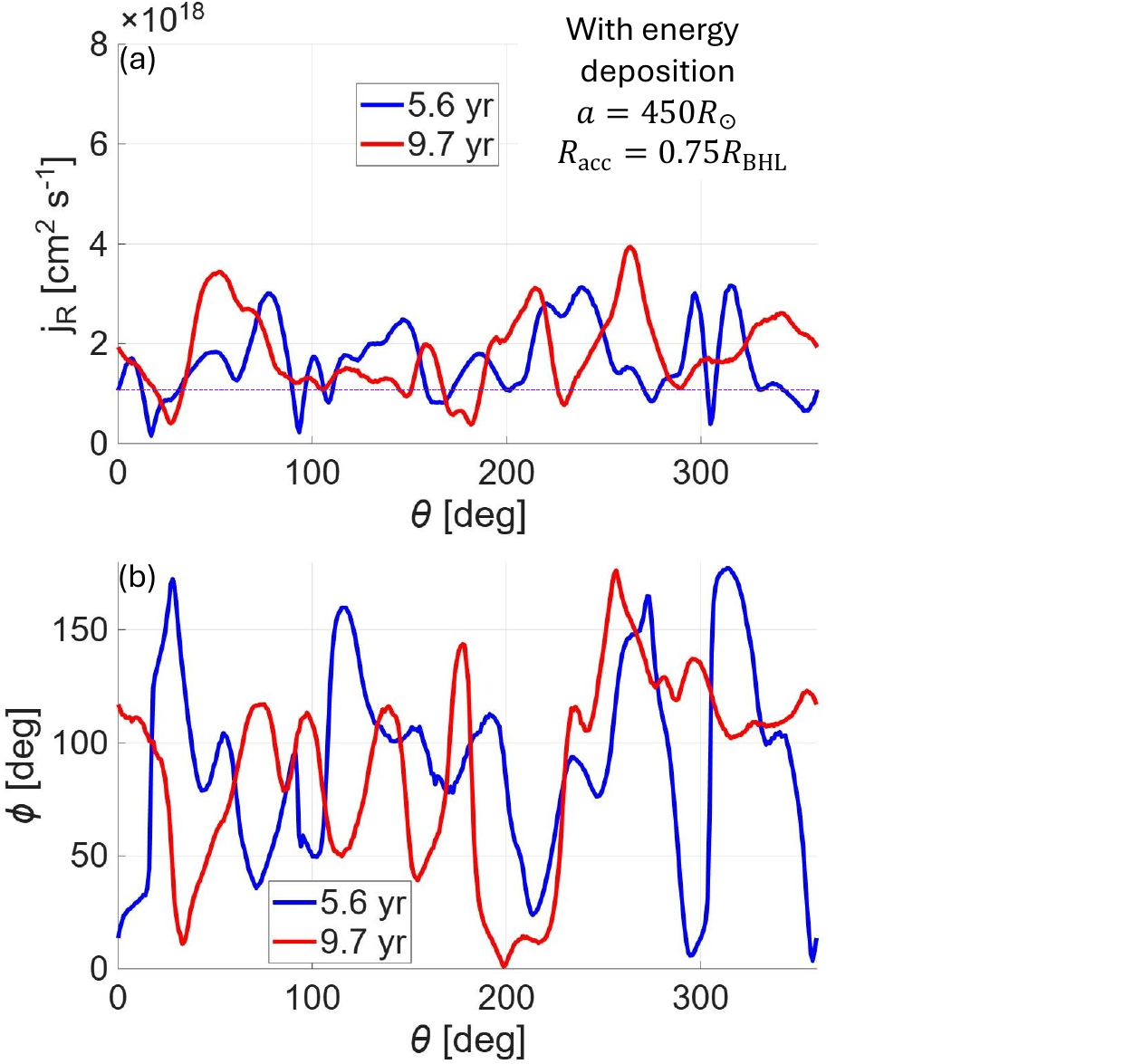}
\caption{Similar to Figure \ref{fig:a700} but for a circular orbit at $a=450 R_\odot$, and only for the simulation with energy deposition.   
}
\label{fig:a450}
\end{figure}

In Figure \ref{fig:a450} we present the results for a circular orbit $a=450 R_\odot$ where $M_1(a)=8.7 M_\odot$, and only for the simulation with energy deposition; in this case $j_{\rm O}(450)=1.1 \times 10^{18} \cm^2 \s^{-1}$, by the scaling of equation (\ref{eq:jaccO2}). The value of $j_{\rm R}$ is lower by a few tens of percent relative to that for $a=700 R_\odot$. The fluctuations are less frequent with respect to the angle: in the case of $a=700 R_\odot$, the direction $\phi$ crosses the angle $90^\circ$ 22 times during one orbit (panel c of Figure \ref{fig:a700}), while in the case of $a=450 R_\odot$ only 10 times. The orbital period at $a=700 R_\odot$ is $P_{\rm orb}(700) = 1.65 \yr$, so that the typical fluctuation period is $\tau_{\rm R} (700) \simeq 0.075 \yr \simeq 27$~day. At $a=450 R_\odot$, $P_{\rm orb}(450) = 0.95 \yr$, and so $\tau_{\rm R} (450) \simeq 0.095 \yr \simeq 35$~day.  

\begin{figure} 
\centering
\includegraphics[trim=0.0cm 9.8cm 0.0cm 0.0cm ,clip, angle=0, scale=0.48]{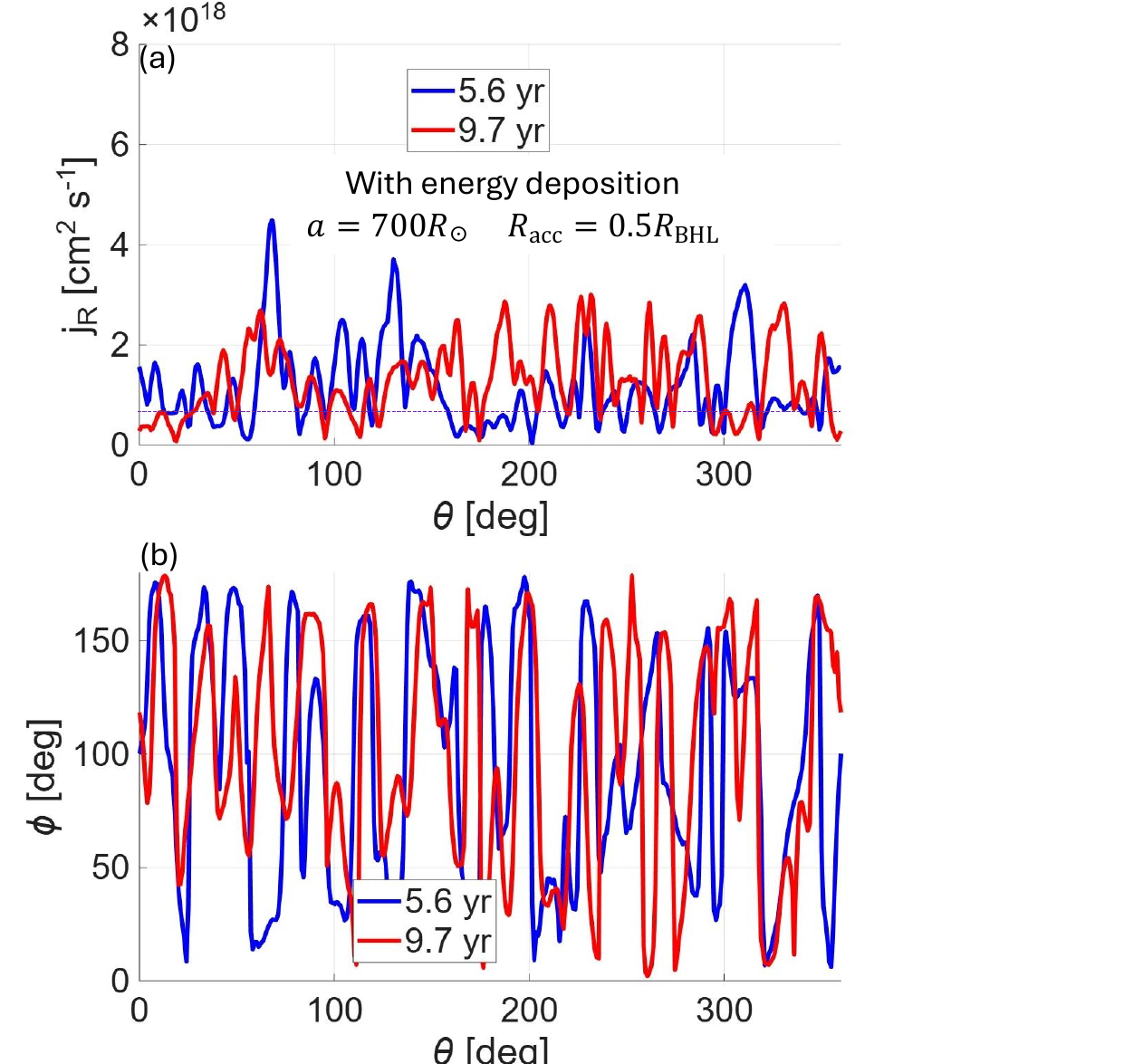}
\caption{Similar to Figure \ref{fig:a700} but for $R_{\rm acc} =0.5 R_{\rm BHL}$, and only the simulation with energy deposition.   
}
\label{fig:a700R05}
\end{figure}

We calculate the case of $R_{\rm acc}= 0.5 R_{\rm BHL}$ instead of $R_{\rm acc}= 0.75 R_{\rm BHL}$ for a circular orbit with $a=700 R_\odot$; we present the results in Figure \ref{fig:a700R05}. The value of $j_{\rm O} \propto R^2_{\rm acc}$ for the same value of the other parameters in equation (\ref{eq:jaccO1}), and so it is $(0.5/0.75)^2=0.44$ times the value for $R_{\rm acc} = 0.75R_{\rm BHL}$, namely, $j_{\rm O}(700,0.5)= 0.64 \times 10^{18} \cm^2 \s^{-1}$ as we show by the purple-dashed horizontal line in panel (a). The typical amplitude of the fluctuating $j_{\rm R}$ in this case (panel a of Figure \ref{fig:a700R05}) is about half that in the $R_{\rm acc}= 0.5 R_{\rm BHL}$ case (panel b of Figure \ref{fig:a700}). Panel (b) of Figure \ref{fig:a700R05} shows that the typical period of fluctuations decreases; the direction $\phi$ crosses the angle $90^\circ$ 31 times (instead of 22 times in panel c of Figure \ref{fig:a700}). This gives a typical fluctuation period of $\tau_{\rm R} (700,0.5) \simeq 0.053 \yr \simeq 19$~day (instead of $\simeq 27$~days). The shorter typical period is due to the shorter crossing time of the convective cells across the smaller accretion radius, and to fewer convective cells inside the accretion radius at each given time (since its volume is smaller).   

\begin{figure} 
\centering
\includegraphics[trim=0.6cm 21.3cm 0.0cm 0.0cm ,clip, angle=0, scale=0.56]{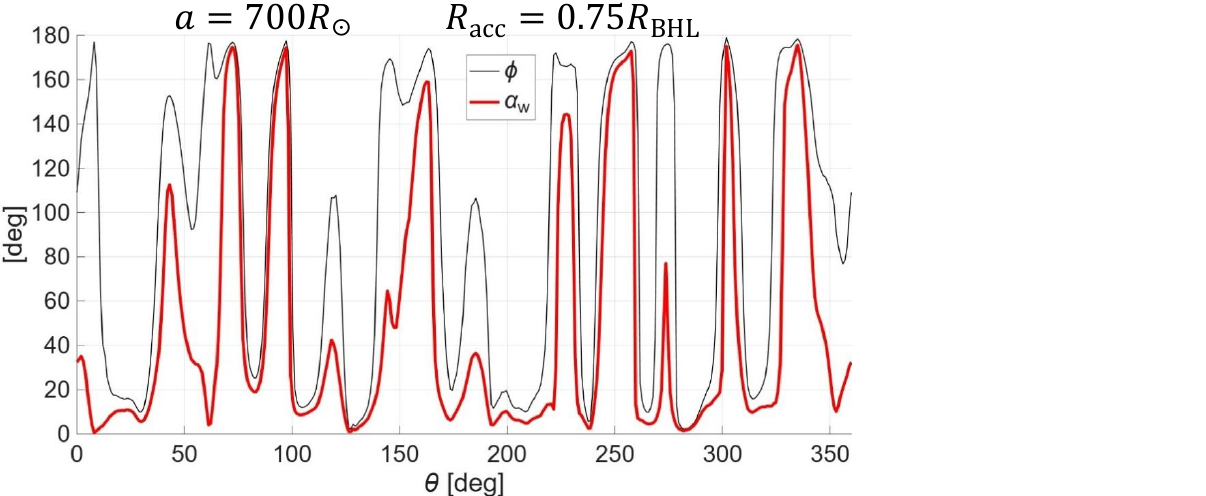}
\caption{The direction of the random component of the accreted angular momentum ($\phi$ by equation \ref{eq:JRRand}; thin black line) and of the total angular momentum ($\alpha_{\rm w}$ by equation \ref{eq:JaccRand}; thick red line) as a function of the orbital angle at $t=1.97 \yr$ for a circular orbit of $a=700 R_\odot$, $R_{\rm acc} =0.75 R_{\rm BHL}$, and for the case with energy deposition.    
}
\label{fig:a700Alpha}
\end{figure}

The lower panels of Figures \ref{fig:a700} - \ref{fig:a700R05} present the angle of the random component of the angular momentum of the accreted mass, $\vec{j}_{\rm R}$, with respect to the orbital angular momentum (equation \ref{eq:JRRand}). Equation (\ref{eq:JaccRand}) gives the direction of the total specific angular momentum of the accreted gas, $\vec{j}_{\rm R}$. In Figure \ref{fig:a700Alpha} we plot these two angles for the case shown in the lower panel of Figure \ref{fig:a700} at $t=1.97 \yr$. In a large fraction of the orbit, the angles are very close in value because the fluctuations are larger than the orbital component; namely, during a large fraction of the orbit $j_{\rm R} \gg j_{\rm O}$. This is evident in panel (b) of Figure \ref{fig:a700} by comparing the solid lines with the dashed-purple horizontal line.     

\section{Discussion and Summary}
\label{sec:Summary}

Jets launched by the more compact companion in CEE and GEE introduce complexity and a wide variety of processes. The wobbling of the jets due to the convective envelope further complicates this. We turn to mention several effects of this wobbling.
Additional, but much smaller, wobbling might occur in the case of a black hole companion due to disk-jet interaction (e.g., \citealt{Gottliebetal2022, Gottliebetal2025}).

We also state again that we consider the standard CEE to be one that includes the jets that the companion is very likely to launch in one, two, or all three phases: (a) just before the CEE, namely, the GEE, (b) during the CEE, or (c) during the exit from the CEE, like accretion from a circumbinary disk. 

\subsection{Summary of results}
\label{subsec:SummaryResults}

Our findings from the simulation with energy deposition, which is a more realistic RSG model than the one without energy deposition (see panel a of Figure \ref{fig:a700}), are that the random component fluctuations, $j_{\rm R}$, can reach a typical amplitude of $\simeq 2-3$ times that of the constant component of the angular momentum.  The specific angular momentum of the combined components can reach values of $j_{\rm acc,max} (700) \simeq 4 \times 10^{18} \cm^2 \s^{-1}$ at an orbit of $a=700 R_\odot$ and an accretion radius of $R_{\rm acc}=0.75 R_{\rm BHL}$ at a large fraction of the orbit (panels b and c of Figure \ref{fig:a700}). The value is about half for $R_{\rm acc}=0.5 R_{\rm BHL}$ at the same radius, i.e., $j_{\rm acc,max} (700,0.5) \simeq 2 \times 10^{18} \cm^2 \s^{-1}$ (Figure \ref{fig:a700R05}). For $a=450 R_\odot$ and an accretion radius of $R_{\rm acc}=0.75 R_{\rm BHL}$, the value is $j_{\rm acc,max} (700) \simeq 3 \times 10^{18} \cm^2 \s^{-1}$ (Figure \ref{fig:a450}).  

We emphasize that in this study, we simulated the RSG model without any influence from the companions, not the orbital energy it deposits in the envelope, not its gravity, and not the jets it might launch. Therefore, our results are approximate and may differ somewhat from those that a full treatment of the CEE will yield. 

\subsection{Comparing with 1D calculations}
\label{subsec:ComparingTo1D}

Since \cite{Dorietal2023} used a 1D model, they could not track fluctuations along an orbit, and their value is an average of those fluctuations. In contrast, they inject the energy the companion injects in the envelope as it spirals in. 
They found the average amplitude of the stochastic angular momentum component to be about 0.1-1 times that of the constant component due to orbital motion. However, they used $\eta=0.2$. For $\eta=0.1$ as we use here (equation \ref{eq:jaccO2}), their results read $(\langle j_{\rm R} \rangle/ j_{\rm O})_{\rm Dori} \simeq 0.2-2$. \cite{Dorietal2023} performed their calculations in 1D using mixing length theory. They also deposited the energy that the companion deposits as it spirals in, so their initial RSG model with a radius of $500 R_\odot$ swelled in some cases. They also removed mass, causing the convective zone to shrink in some cases. Considering that they obtained average values, and the large differences between the 1D RSG model they used and the 3D model we used here, the results of the average (typical) fluctuations of $j_{\rm R}$ are similar.   
However, while the average of the fluctuations of $j_{\rm R}$ causes a typical change in the direction of the angular momentum of the accreted mass by tens of degrees, but generally not a full flip, we find that some fluctuations flip the angular momentum direction of the accreted mass (Figure \ref{fig:a700Alpha}).  
  
\subsection{Implications to the accretion disk}
\label{subsec:AccretionDisk}

The Keplerian orbit of the accreted mass is $R_{\rm d2} = 1.2 (j_{\rm acc}/4 \times 10^{18} \cm^2)^2(M_2/1.4 M_\odot)^{-1} R_\odot$ (equation \ref{eq:Rdisk2}). Accretion disks easily form around NSs and black holes, but marginally around main-sequence and Wolf-Rayet type stars (with radii comparable to or larger than the value of $R_{\rm d2})$, and possibly only during short times of large $j_{\rm R}$ fluctuations. For an NS inside an RSG envelope, which accretes mass and launches jets, namely, a CEJSN impostor (common-envelope jets supernova impostor), the wobbling jets imply that they will mostly be choked inside the envelope. We expect no constant-axis jets. This might change when the NS enters the core, where some zones might have no convection and most of the angular momentum of the accreted mass comes from orbital motion as the NS enters the core and launches jets in a CEJSN event.   

Jets launched by main-sequence stars can influence the onset and early phases of the CEE, as the main-sequence companion enters the RSG (or AGB, or RGB) envelope with an already-existing, developed accretion disk which was formed by Roche-Lobe overflow. Particularly, a relatively massive main-sequence companion can experience the GEE, as their jets might remove a large amount of RSG envelope mass as the companion star grazes the envelope (for some GEE simulations, see, e.g., \citealt{Shiber2018Galax, LopezCamaraetal2022, ShiberIaconi2024}). 
   
As in \cite{Dorietal2023}, we take the accretion disk lifetime (without fresh mass supply), which is about its viscous timescale or relaxation time, to be $\zeta \simeq 100$ times longer than the Keplerian orbit. For $M_2=1.4 M_\odot$ and $R_{\rm d2}=1 R_\odot$, the lifetime of the disk by viscous dissipation is $\tau_{\rm d2} \simeq 10(\zeta/100)$~day. The large $j_{\rm R}$ fluctuations last for $\simeq 5-10$ days. Therefore, the outer part of the accretion disk might have no time to relax. The inner disk around an NS or a black hole has much shorter timescales, and such accretion disks have time to relax and launch jets.  
   
Accretion disks around main-sequence stars, when they form, will have no time to relax between angular momentum fluctuations in CEE. If the disk launches jets, the two opposite jets might be unequal, as the accretion disk has no time to relax. The formation of accretion disks around main-sequence companions in the GEE and the CEE should be a high-priority topic of future studies, particularly considering the effect of jets on the mass-accreting companion. The main effect of the jets on the mass-accreting companion might be the removal of high-entropy accreted gas from its envelope outskirts, a process that facilitates further accretion and jet-launching; this process is termed ``jetted mass removal accretion scenario'' (e.g., \citealt{CohenBearSoker2026}).

\subsection{Implications to the jet feedback mechanism}
\label{subsec:Feedback}

The jets that the companion launches in the CEE or the GEE operate in a negative feedback mechanism. Namely, the jets remove mass and deposit energy that swell the envelope, effects that reduce the density in the companion vicinity, hence reducing the accretion rate and jet power (e.g., \citealt{GrichenerCohenSoker2021, Hilleletal2022, Gurjaletal2026}). There is also a positive effect, by which the jets remove energy and gas from the very close vicinity of the accreting companion, hence allowing accretion to proceed.  
 
The stochastic component of the angular momentum of the accreted gas contributes to the negative feedback mechanism. The wobbling jets imply that the jets are launched in different directions, not only perpendicular to the orbital plane. In many cases, one jet will be launched towards the center of the giant star, heating regions the companion has not yet reached. The opposite jet will more easily break out from the envelope and add to the clumpy structure of the ejected mass. The energy that the jets inject into the envelope influences the convection. Fully simulating the feedback with the new RSG stellar model we constructed will require additional computing resources, but we aim to conduct this study.   

\subsection{Shaping the ejecta}
\label{subsec:Ejecta}
The orbital motion of the jet-launching companion implies that the jets have a hard time drilling through the envelope even without the random angular momentum component  (e.g., \citealt{Papishetal2015, LopezCamaraetal2022}).
Our finding of large-amplitude wobbling, as \cite{Dorietal2023} noted, implies that, in general, jets are less likely to penetrate the envelope. 
However, the large fluctuations we found here imply that, in many cases, one jet will be launched radially outward; as noted above, such jets might break out from the envelope in a plane near the equatorial plane. 

The wobbling jets that break in different directions might inflate medium-size bubbles in the ejecta, which in turn can compress shells on their outskirts. 
These will be of a medium size, larger than small-scale instabilities and smaller than large bipolar lobes. Such shells form filaments and arcs in the ejecta, and later in the observed nebula. It is possible that some arcs and filaments in planetary nebulae are due to wobbling jets from a main-sequence companion in an AGB envelope.

\section*{Acknowledgments}

We thank an anonymous referee for helpful comments, particularly regarding the figures. 
A grant from the Pazy Foundation (2026) supported this research.

\bibliography{bib}{}
\bibliographystyle{aasjournalv7.1}
\end{document}